\documentclass[a4paper,12pt,useAMS]{emulateapj}
\usepackage{txfonts}
\usepackage{amssymb}

\usepackage{graphicx,graphics}
\usepackage{natbib}
\def\km{\,{\rm km}}
\def\cm{\,\rm cm}
\def\s{\,{\rm s}}

\def\mpc{\,{\rm Mpc}}
\def\msun{\,{\rm M_{\odot}}}
\def\erg{\,{\rm erg}}
\def\kev{\,{\rm keV}}

\def\yr{\,{\rm  yr}}

\def\Gpc{\,{\rm Gpc}}

\begin{document}
\title{Constraining warm dark matter mass with cosmic reionization and gravitational wave}
\author{Wei-Wei Tan$^{1,2}$, F. Y. Wang$^{1,2}$ \& K. S. Cheng$^3$}

\affil{$^1$School of Astronomy and Space Science, Nanjing University, Nanjing 210093, China, fayinwang@nju.edu.cn\\
$^2$Key laboratory of Modern Astronomy and Astrophysics (Nanjing
University), Nanjing 210093, China\\
$^3$Department of Physics, University of Hong Kong, Pokfulam Road,
Hong Kong, China}

\begin{abstract}
We constrain the warm dark matter (WDM) particle mass with the
observations of cosmic reionization and CMB optical depth. We
suggest that the GWs from stellar mass black holes (BHs) could give
a further constraint on WDM particle mass  for future observations.
The star formation rates (SFRs) of Population I/II (Pop I/II) and
Population III (Pop III) stars are also derived. If the metallicity
of the universe have been enriched beyond the critical value of
$Z_{\rm crit}=10^{-3.5}Z_{\odot}$, the star formation shift from Pop
III to Pop I/II stars. Our results show that the SFRs are quite
dependent on the WDM particle mass, especially at high redshifts.
Combing with the reionization history and CMB optical depth derived
from the recent \emph{Planck} mission, we find that the current data
requires the WDM particle mass in a narrow range of $1\kev \lesssim
m_{\rm x}\lesssim 3\kev$. Furthermore, we suggest that the
stochastic gravitational wave background (SGWB) produced by stellar
BHs could give a further constraint on the WDM particle mass for
future observations. For $m_{\rm x}=3 \kev$ with Salpeter (Chabrier)
initial mass function (IMF), the SGWB from Pop I/II BHs has a peak
amplitude of $\Omega_{\rm GW}\approx2.8\times 10^{-9}~(5.0\times
10^{-9})$ at $f= 316 {\rm Hz}$, while the GW radiation at $f<10$Hz
is seriously suppressed. For $m_{\rm x}=1 \kev$, the SGWB peak
amplitude is the same as that of $m_{\rm x}=1\kev$, but a little
lower at low frequencies. Therefore, it is hard to constrain the WDM
particle mass by the SGWB from Pop I/II BHs. To assess the
detectability of GW signal, we also calculate the signal to noise
ratio (SNR), which are $\rm SNR=37.7~ (66.5)$ and $27~(47.7)$ for
$m_{\rm x}=3\kev$ and $m_{\rm x}=1\kev$ for Einstein Telescope (ET)
with Salpeter (Chabrier) IMF, respectively. The SGWB from Pop III
BHs is seriously dependent on the WDM particle mass, the GW strength
could be an order of magnitude different and the frequency band
could be two times different for $m_{\rm x}=1\kev$ and $m_{\rm
x}=3\kev$. Moreover, the SGWB from Pop III BHs with $m_{\rm
x}=1\kev$ could be detected by LISA for one year of observation, but
can not for $m_{\rm x}=3\kev$.
\end{abstract}

\keywords{cosmology: theory - dark matter - cosmology: reionization
- gravitational waves}


\section{Introduction}

The astrophysical and cosmological probes have confirmed that
baryons constitute only some $16\%$ of the total matter in the
Universe. The rest of the mass is in the form of `dark matter' (DM).
The nature of DM particles is poorly understood, as they do not
interact with baryons. Many indirect searches have been carried out,
including searching for $\gamma$-ray signals at the Galactic center,
in nearby galaxies, and the diffuse $\gamma$-ray background
\citep{Ackermann12, Ackermann14, The Fermi LAT collaboration15}.
However, none of them could provide robust evidence for the
observation of DM. The GeV $\gamma$-ray excess from the Galactic
center could be a signal of DM annihilation, but still can not be
confirmed \citep{Daylan15, Zhou15}.

Among various DM candidates, the most popular candidate is the
weakly interacting massive particles (WIMPs; like the neutralino),
which have mass in GeV range \citep{Jungman96, Bertone05, Hooper07,
Feng10}. The WIMPs are non-relativistic at the epoch of decoupling
from the interacting particles and have negligible free-streaming
velocities. Therefore, they are `cold', called cold dark matter
(CDM). In CDM scenario, `halos' formed in small clumps, and then
merged together into larger and massive objects. Galaxies formed in
these halos are because of the cooling of atomic hydrogen
\citep[H;][]{Tegmark97} or molecular hydrogen \citep[${\rm
H_2}$;][]{Ciardi00, Haiman00}. On large cosmological scales (from
the range $\sim1\Gpc$ down to $\sim10\mpc$), CDM paradigm has great
success in explaining the observed universe and reproducing the
luminous structures \citep{Fixsen96, Borgani01, Lange01, Cole05,
Tegmark06, Benson10, Wang13, Hinshaw13, Slosar13, Planck
Collaboration14, Wei16}. However, on small scales ($\lesssim1\mpc$),
there are still some discrepancies between the CDM paradigm and
observations: (a) the core-cusp problem \citep{Navarro97,
Subramanian00}. CDM simulations predict a cusp-core DM halo, whereas
the observations find them cored \citep{Salucci12}; (b) too big to
fail problem \citep{Boylan-Kolchin12}. CDM simulations predict a
central DM density significantly higher than the observation that
allowed; and (c) the `missing satellite problem'. N-body simulations
based on the CDM paradigm predict a number of subhalos larger than
that of satellites found in our Galaxy \citep{Klypin99, Moore99,
Papastergis11}. Many methods have been proposed to solve these small
scale problems, such as modifying the nature of DM from the CDM
paradigm \citep{Hu00, Spergel00, Su11, Menci12}, adding supernova
feedback effect in simulation \citep{Weinberg02, Mashchenko06,
Governato10, Pontzen14}, and considering the interplay between DM
and baryons during the formation of the galaxy \citep{El-Zant01,
Tonini06, Pontzen14}. However, these methods are insufficient to
solve all the above problems.

Alternatively, a more possible solution to these small scale
problems is the warm dark matter (WDM) scenario, with DM particle
mass in $\kev$ range. The candidates are sterile neutrinos
\citep{Dodelson94, Abazajian01, Abazajian06, Shaposhnikov06,
Boyarsky09, Kusenko09, Abazajian12} and gravitinos
\citep{Kawasaki97, Gorbunov08}. WDM particles are lighter than CDM
particles, so they could remain relativistic for longer time in the
early universe and retain a non-negligible velocity dispersion. They
are more easy to free-stream out from small scale perturbations, and
suppress the formation of subhalos \citep{Bode01, Lovell14}. The
most powerful test for WDM scenario is the high-redshift universe. A
number of works have been done to constrain the WDM particle mass
($m_{\rm x}$). For example, \cite{Kang13} gave a lower limit of
$m_{\rm x}\gtrsim 0.75\kev$ by reproducing the stellar mass
functions and Tully-Fisher relation for $0<z<3.5$ galaxies.
\cite{Viel13} used Lyman-$\alpha$ flux power spectrum measured from
high-resolution spectra of 25 quasars to obtain a lower limit of
$m_{\rm x}\gtrsim 3.3\kev$. \cite{de Souza13} used high-redshift
($z>4$) gamma-ray bursts to constrain $m_{\rm x}\gtrsim 1.6-1.8
\kev$. \cite{Dayal15b} constrained $m_{\rm x}\gtrsim 2.5 \kev$ by
comparing the semi-analytic merger tree based framework for
high-redshift ($z\simeq 5-20$) galaxy formation with reionization
indicators. \cite{Lapi15} gave a narrow constraint of $2<m_{\rm x}<3
\kev$ by combining the measurements of the galaxy luminosity
functions out $z\sim 10$ from \emph{Hubble Space Telescope} (HST)
with the reionization history of the universe from the \emph{Planck}
mission. \cite{Pacucci13} constrained $m_{\rm x}\gtrsim 1 \kev$ by
using the number density of $z\approx 10$ lensed galaxies.

Given that structures are formed hierarchically and WDM scenario
smears out the power on small scale, the number density of the
smallest halos (or galaxies) at high redshift will be strongly
decreased, and then the SFR. Especially the SFRs of Pop III stars
and high-redshift Pop I/II stars, because they are firstly formed in
these small halos \citep{Barkana01}. Pop III stars are the massive
stars with masses $\gtrsim 100 M_\odot$ \citep[e.g.,][]{Bromm99,
Abel00, Nakamura01}, which are formed in metal-free gas. The deaths
of Pop III stars lead to the metal enrichment of intergalactic
medium (IGM) via supernova feedback, and subsequently the formation
of Pop I/II stars \citep[the cricitical metallicity is
$10^{-3.5}Z_{\rm \odot}$;][]{Ostriker96, Madau01, Bromm03,
Furlanetto03}. The mass of Pop I/II stars is in the range of
$0.1\sim 100 M_{\odot}$. The first light from Pop III stars brought
the end of the cosmic dark ages, and then the universe began to
reionize. Recent observation from \emph{Planck} mission measured the
integrated CMB optical depth with $\tau=0.066^{+0.013}_{-0.013}$
\citep[with the constraint from \emph{Planck} TT+low
polarization+lesing+BAO;][]{Planck Collaboration15}, and most of the
observations show that the universe was fully reionized at redshift
$z\simeq 6$ \citep{Chornock13, Treu13, Pentericci14, Schenker14,
McGreer15}. These measurements gauge the level of the reionization
history from the high-redshift stars. Furthermore, as the
high-redshift stars formed in small halos are greatly affected by
the halo number density, their formation rates could provide a
indirect test on the WDM scenario.

On September 14, 2015 the Advanced LIGO observed the
gravitational-wave event GW150914 \citep{Abbott16a}. The observed
signal is consistent with a black-hole binary waveform with the
component masses of $m_1=36^{+5}_{-4}\msun$ and
$m_2=29^{+4}_{-4}\msun$, which demonstrates the existence of
stellar-mass black holes massive than $25 \msun$. The second GW
candidate GW 151226 was observed by the twin detectors of the
Advanced LIGO on December 26, 2015 \citep{Abbott16b}. The inferred
initial BH masses are $14.2_{-3.7}^{+8.3}\msun$ and
$7.5_{-2.3}^{+2.3}\msun$, and the final BH mass is
$20.8_{-1.7}^{+6.1}\msun$. The decay of the waveform at the final
period are also observed, which are consistent with the damped
oscillations of a black hole relaxing to a stationary Kerr
configuration. The collapses of Pop I/II or Pop III stars into black
holes (BHs) could also release gravitational waves
\citep[GWs;~][]{Buonanno05, Sandick06, Suwa07, Pereira10, Ott13,
Yang15}, which is dominated by `quasi-normal ringing' of a perturbed
black hole. Therefore, it is expected that the Advanced LIGO could
also observe this kind of gravitational wave radiations. In this
paper, we will calculate the SGWB from BH `ringing', which relates
with the de-excitation of the BH quasi-normal modes. Because the
SGWB is quite dependent on the SFR, it could be used to constrain
the WDM particle mass indirectly. Several GW detectors are operating
or planed in future: advanced VIRGO and LIGO working at $\approx\rm
10 Hz-3kHz$, the Einstein Telescope (ET) with the sensitive
frequency of $1-100 \rm Hz$, the Laser Interferometer Space Antenna
\footnote{http://lisa.nasa.gov/} (LISA) covering the frequency range
of $10^{-4}-0.1 \rm Hz$, the Decihertz Interferometer Gravitational
wave
Observatory\footnote{http://universe.nasa.gov/program/vision.html}
(DECIGO)\citep{Kudoh06}, and the Big Bang Observer (BBO) operating
in the range $0.01-10$ Hz. Therefore, GW signal from BHs ringing
will open a new window for the restriction of the WDM particle mass.

This paper is organized as follows. In section 2, we describe the
hierarchical formation scenario in the framework of WDM paradigm. In
section 3, we construct the SFRs of Pop I/II and Pop III stars, and
compare them with the recent observations. In section 4, we
constrain the WDM particle mass with the CMB optical depth and the
reionization history. In section 5, we calculate the SGWBs from Pop
I/II and Pop III BHs. Finally, conclusion and discussion are given
in section 6. Throughout this paper, we adopt the standard flat
cosmology with cosmological parameters $\Omega_\Lambda=0.72,
\Omega_{\rm m}=0.28, \Omega_{\rm b}=0.046,
H_0=70\km\s^{-1}\mpc^{-1}$, and $\sigma_8=0.82$.

\section{Hierarchical formation scenario in WDM model}

In the framework of hierarchical formation scenario, \cite{Press74}
first gave a straightforward semi-analytic approach for the
abundance of dark matter halos, which is known as the
Press-Schechter (PS) formalism. An improved PS-like simulation was
proposed by \cite{Sheth99}, and they considered the collapse of the
ellipsoidal halo rather than the spherical one. In quantitative
studies \citep{Greif06, Heitmann06, Reed07}, the Sheth-Tormen (ST)
approach was proved to be more accurate. Therefore, we also choose
ST formalism in our following calculations.

Based on the ST formalism, the halo mass function could be described
as
\begin{equation}
f_{\rm ST}({\sigma}) = A\sqrt{\frac{2a_1}{\pi}}
\left[1+\left(\frac{\sigma^2}{a_1\delta_{\rm c}^2}\right)^p\right]
\frac{\delta_{\rm c}}{\sigma}\exp{\left[-\frac{a_1\delta_{\rm
c}^2}{2\sigma^2}\right]},
\end{equation}
where $A=0.3222, a_1 = 0.707, p = 0.3$ is the best-fitting values
from simulations, $\delta_{\rm c} = 1.686$ is the critical over
density, and $\sigma(M,z)$ is the variance of the linear density
field. The number of dark matter halos per comoving volume at a
given redshift within the mass interval $M\sim M+{\rm d}M$ could be
related to $f_{\rm ST}$ as
\begin{equation}
{\rm d}n_{\rm ST}(M,z)={\rho_{\rm m}\over M} {{\rm d}{\rm
ln}\sigma^{-1}\over {\rm d}M} f_{\rm ST}({\sigma}) {\rm d}M,
\end{equation}
where $\rho_{\rm m}$ is the mean density of the universe. In a
Gaussian density field, the variance of the linear density field in
the local universe with mass $\rm M$ is given by
\begin{equation}
\sigma^2(M, 0)={1\over 2\pi^2}\int^{\infty}_0 k^2 P(k)W^2(k M){\rm
d}k,
\end{equation}
where $P(k)$ is the power spectrum of the density fluctuations at
$z=0$. $W(k, M)$ is the sharp \emph{k}-space filtering with
\begin{equation}
W(k,M)=\\
\left\{ \begin{array}{lcl} 1, & {\rm if}~
k\leq k_s(R),\,\\
0, & {\rm if}~ k> k_s(R),\,
\end{array}\right. \label{sharp-k}
\end{equation}
where $R=(3M/4\pi\rho)^{1/3}$ is the radius of the halo with mass
$M$, and $k_{\rm s}=a/R$ with $a=2.5$ \citep{Benson13}. The redshift
dependence enters only through the linear growth factor $D(z)$,
which could be taken as $D(z)=g(z)/[g(0)(1+z)]$ with
\begin{equation}
g(z)\approx{5 \Omega_{\rm m}(z)\over 2[\Omega_{\rm
m}(z)^{4/7}-\Omega_{\Lambda}(z)+(1+{\Omega_{\rm
m}(z)\over2})(1+{\Omega_{\Lambda}(z)\over70})]},
\end{equation}
therefore, $\sigma(M,z)=\sigma(M, 0)D(z)$.

In the CDM model, the primordial power spectrum is assumed to be
power dependent on scale and multiplied by a transfer function,
where the fluctuations are only determined by the interplay between
self-gravitation, pressure and damping processes. However, in the
WDM model, the WDM particles are relativistic. So the linear
fluctuation amplitude is suppressed below the free-streaming scale
of the WDM particle. The comoving free-streaming scale is given by
\citep{Bode01}
\begin{equation}
\lambda_{\rm fs}\approx 0.11 \left({\Omega_{\rm x} h^2\over
0.15}\right)^{1/3}\left({m_{\rm x}\over\kev}\right)^{-4/3}\mpc,
\end{equation}
where $\Omega_{\rm x}$ is the fraction of the energy density in WDM
particles relative to the critical energy density, $h$ is the Hubble
constant in unites of $100 \km\s^{-1}\mpc^{-1}$, and $m_{\rm x}$ is
the WDM particle mass. Following \cite{Bode01}, the power spectrum
should be modified by a transfer function below the free-streaming
scale, which can be described by
\begin{equation}
P_{\rm WDM}(k) = P_{\rm CDM}(k)\left[1+(\epsilon
k)^{2\mu}\right]^{-5\mu},
\end{equation}
where $\mu=1.12$, $P_{\rm WDM}(k)$ and $P_{\rm CDM}(k)$ are the
power spectra in WDM and CDM paradigms, respectively. $\epsilon$ is
related with both the WDM particle mass $m_{\rm x}$ and the energy
density fraction $\Omega_{\rm x}$, which can be written as
\begin{equation}
\epsilon = 0.049 \left(\frac{\Omega_{\rm
x}}{0.25}\right)^{0.11}\left(\frac{m_{\rm x}}{\rm
keV}\right)^{-1.11}\left(\frac{h}{0.7}\right)^{1.22}h^{-1} \rm Mpc.
\end{equation}

As mentioned above, the smallest structure formed in WDM model is
greatly suppressed by the residual velocity dispersion of WDM
particles. Therefore, the minimum halo mass should also be quite
dependent on the WDM particle mass. We describe the minimum halo
mass as \citep{de Souza13}
\begin{equation}
M_{\rm WDM} \approx 1.8\times10^{10}\left(\frac{\Omega_{\rm
x}h^2}{0.15}\right)^{1/2}\left(\frac{m_{\rm x}}{1 \rm
keV}\right)^{-4}\msun  .
\end{equation}
Then, we could calculate the fraction of the baryons inside
structures. Here we assume that the baryon distribution traces the
dark matter distribution without bias, and the baryonic density is
proportional to the density of dark matter. Therefore, the baryonic
fraction in structures can be given by
\begin{equation}
f_{\rm b}(z)={\int_{M_{\rm min}}^{\infty} n_{\rm ST}(M,z) M {\rm d}M
\over \rho_{\rm m}},\label{fraction b}
\end{equation}
where $\rho_{\rm m}$ is the mean density of the universe.
Considering that the minimum halo should be capable of forming
stars, we rewrite the minimum halo mass as
\begin{equation}
 M_{\rm min} = {\rm Max}[M_{\rm gal}(z), M_{\rm WDM}(m_{\rm x})],
\end{equation}
where $M_{\rm gal}(z)$ corresponds to the halo mass that could be
efficiently cooling by ${\rm H_2}$ gas with the virial temperature
$T_{\rm vir}=10^4$ Kelvin, which could be given by
\begin{equation}
 M_{\rm gal}(z) \approx 10^8 \times \left({\eta\over 0.6}\right)^{-2/3}
 \left({T_{\rm vir}\over 10^4 {\rm K}}\right)^{3/2}\left({1+z\over 10}\right)^{-3/2}\msun,
\end{equation}
here $\eta=1.22$ is the mean molecular weight. In fact, $M_{\rm
WDM}>M_{\rm gal}$ for $m_{\rm x}<2\kev$, and hence $M_{\rm min}$ is
not sensitive to the exact value of $M_{\rm gal}$ for low WDM
particle mass.

The baryonic fraction in structures can be given by equation
(\ref{fraction b}), therefore, we could describe the accretion rate
of baryon into structures at different cosmic time. Following
\cite{Daigne06}, it could be described as
\begin{equation}
a_{\rm b}(t) = \Omega_{\rm b}\rho_{\rm c}\left(\frac{{\rm d}t}{{\rm
d}z}\right)^{-1}\left|\frac{{\rm d}f_{\rm b}(z)}{{\rm d}z}\right|
\label{accreation rate},
\end{equation}
where $\rho_{\rm c}$ is the critical density of the universe. The
age of the universe could be related to the redshift by
\begin{equation}
\frac{{\rm d}t}{{\rm d}z} = \frac{9.78h^{-1}
\rm{Gyr}}{(1+z)\sqrt{\Omega_{\Lambda}+\Omega_{\rm
m}(1+z)^{3}}}.\label{timez}
\end{equation}

Thereafter, we will calculate the formation rates of Pop I/II and
Pop III stars in the framework of WDM paradigm, and study the effect
of the WDM particle mass on their formation rates.

\section{Cosmic star formation rate}

First, we should make it clear that how the matter transfer among
stars, interstellar medium (ISM) and intergalactic medium (IGM).
Four fundamental processes should be included: (a) the accretion of
baryons from IGM to form structures, $a_{\rm b}(t)$; (b) the
transfer of baryons from structures (or ISM) into stars, $\Psi(t)$;
(c) stars return mass to the ISM through stellar winds and
supernovae, $M_{\rm ej}(t)$; (d) the outflow of baryons from
structures into IGM through galactic winds and direct ejecta of
stellar supernova, $o(t)$. In this section, we will describe how to
calculate the above four processes. For the first point, the
baryonic accretion rate $a_{\rm b}(t)$ is described by equation
(\ref{accreation rate}). For the second point, we will calculate
both the formation rates of Pop I/II and Pop III stars. For
simplicity, we assume both the SFRs follow the Schmidt law
\citep{Schmidt59, Schmidt63}, i.e., SFRs are proportional to the gas
density $\rho_{g}(t)$ in structures. Here we employ an exponentially
decreasing SFR for Pop I/II stars, which fits the observational data
quite well \citep{Daigne06}. Hence, we take

\begin{equation}
\Psi_{\rm I/II}(t)=f_1 {\rho_{\rm g}(t)\over \tau_1}e^{-(t-t_{\rm
init})/\tau_1}(1-e^{-Z_{\rm IGM}/Z_{\rm crit}}),
\end{equation}
where $t_{\rm init}$ is the initial age of the universe at redshift
$z_{\rm init}$ and $\tau_1$ is the star formation time scale. Here
we consider $f_1 e^{-(t-t_{\rm init})/\tau_1}$ as the star formation
efficiency of Pop I/II stars, which could be determinated by the
observed SFR at low redshift. The last term represents the fraction
of gas that are metal populated by outflows of structures, where
$Z_{\rm IGM}$ is the metallicity of IGM and $Z_{\rm
crit}=10^{-3.5}Z_{\odot}$ is the critical metallicity. For Pop III
stars, we assume an exponential decease star formation model, which
could be described as (Daigne 2006)
\begin{equation}
\Psi_{\rm III}(t)=f_2 {\rho_{\rm g}(t)}e^{-Z_{\rm IGM}/Z_{\rm
crit}}.
\end{equation}
For the star formation efficiency, we set a typical value of
$f_2=4.5\%$ \citep{Daigne04}.  Such a value is just located within
its theoretically-expected range of $\sim 10^{-6}-10^{-3}$
\citep{Greif06, Marassi09}. Therefore, the total mass rate that goes
into stars is
\begin{equation}
\frac{{\rm d}^{2}M_{\star}}{{\rm d}V{\rm d}t} =\Psi_{\rm
I/II}(t)+\Psi_{\rm III}(t)\label{sfr},
\end{equation}
which is the sum formation rate of Pop I/II and Pop III stars.

We assume two IMFs for both Pop I/II and Pop III stars. The Salpeter
IMF \citep[SIMF;][]{Salpeter55} is
\begin{equation}
\Phi(m)\propto m^{-(1+x)},\label{Salpeter}
\end{equation}
with $x=1.35$. The Chabrier IMF \citep[CIMF;][]{Chabrier03} is
\begin{equation}
\Phi(m)\propto \\
\left\{ \begin{array}{lcl} {0.158\over m} {\rm exp}{[-{\rm
log}(m/\msun)-{\rm log}(0.08)]^2\over 2\times (0.69)^2}, & {\rm if}~
m\leq 1 \msun,\,\\
m^{-2.3}, & {\rm if}~ m> 1 \msun,\,
\end{array}\right. \label{Chabrier}
\end{equation}
The two IMFs are normalized independently for these two kind of
stars, but with different integrate mass range. We consider $m_{\rm
inf}=0.1{\rm M}_{\odot}$ and $m_{\rm sup}=100{\rm M}_{\odot}$ for
Pop I/II stars, $m'_{\rm inf}=100{\rm M}_{\odot}$ and $m'_{\rm
sup}=500{\rm M}_{\odot}$ for Pop III stars.

For the third point, the mass ejected from stars through stellar
winds and supernovae into ISM is given by
\begin{eqnarray}
\frac{{\rm d}^{2} M_{\rm ej}}{{\rm d}V{\rm d}t} &=& \int_{m(t)}^{\rm
M_{sup}}{(m-m_{\rm
r})\Phi(m)\Psi_{\rm I/II}(t-\tau_{\rm m}){\rm d}m}+\nonumber\\
&&\int_{m'(t)}^{\rm M'_{sup}}{(m'-m'_{\rm r})\Phi(m')\Psi_{\rm
III}(t-\tau_{\rm m'}){\rm d}m'},\label{mej}
\end{eqnarray}
where $m(t)$ corresponds to the stellar mass whose lifetime is equal
to the age of the universe ($t$). We use the mass-lifetime relation
proposed by \cite{Scalo86} and \cite{Copi97} to derive $m(t)$ or the
stellar lifetime $\tau_{\rm m}$. The mass remnant $m_{\rm r}$
depends on the mass of progenitor, and we give a description for
$m_{\rm r}$ as follows \citep{Pereira10}:

a) Stars with $m < 1\ {\rm M}_{\odot}$ do not contribute to $M_{\rm
ej}$;

b) Stars with $1\ {\rm M}_{\odot} \leq m\leq 8\ {\rm M}_{\odot}$
dies as the carbon-oxygen white dwarfs with remnants

\begin{equation}
m_{\rm r} = 0.1156\ m +0.4551;
\end{equation}

c) Stars with $8\ {\rm M}_{\odot} < m\leq 10\ {\rm M}_{\odot}$ left
the oxygen-neon-magnesium white dwarfs as the remnants with $m_{\rm
r} = 1.35\ {\rm M}_{\odot}$;

d) Stars with $10\ {\rm M}_{\odot} < m< 25\ {\rm M}_{\odot}$ left
neutron stars as remnants ($m_{\rm r}=1.4\ {\rm M}_{\odot}$);

e) Stars with $25\ {\rm M}_{\odot} \leq m\leq 140\ {\rm M}_{\odot}$
produce black hole remnants equal to the helium core before collapse
with (see Heger \& Woosley 2002),

\begin{equation}
m_{\rm r}=m_{\rm He}=\frac{13}{24}(m-20\ {\rm M}_{\sun}).
\end{equation}

f) Stars with $140\ {\rm M}_{\odot} \leq m\leq 260\ {\rm M}_{\odot}$
explode as pair-instability supernova (PISN) (Heger \& Woosley
2002)) and left nothing.

g) Stars with $260\ {\rm M}_{\odot} \leq m\leq 500\ {\rm M}_{\odot}$
collapse directly into BHs without mass lose.

For the fourth point, the outflow of baryons from structures into
the IGM could be computed by
\begin{eqnarray}
&o(t)={2 \epsilon \over v_{\rm esc}^2(z)}\int_{{\rm max}[8\msun,
m_{\rm d}(t)]}^{m_{\rm sup}}{\rm d}m \Phi(m) \Psi_{\rm
I/II}[t-\tau(m)]E_{\rm kin}(m)+\nonumber\\
&{2 \epsilon \over v_{\rm esc}^2(z)}\int_{{\rm max}[100\msun,
m'_{\rm d}(t)]}^{m'_{\rm sup}}{\rm d}m' \Phi(m') \Psi_{\rm
III}[t-\tau(m')]E'_{\rm kin}(m'),\label{Outflow}
\end{eqnarray}
where $E_{\rm kin}$ is the kinetic energy released by the explosion
of a star with mass $m$, we give a value of $E_{\rm
kin}=10^{51}\erg$ for Pop I/II stars and $E'_{\rm kin}=10^{52}\erg$
for Pop III stars. $\epsilon=10^{-3}$ is the fraction of kinetic
energy that is available to power the outflow. $v^2_{\rm esc}(z)$ is
the mean square of the escape velocity of structures at redshift z
\citep[e.g.,][]{Scully97}, which cold be described by
\begin{eqnarray}
v_{\rm esc}^2(z)={\int_{M_{\rm min}}^\infty{\rm d}M n_{\rm
ST}(M,z)M(2GM/R)\over \int_{M_{\rm min}}^\infty{\rm d}M n_{\rm
ST}(M,z)M}
\end{eqnarray}

Combing equations (\ref{accreation rate}), (\ref{sfr}), (\ref{mej})
and (\ref{Outflow}), we could derive the evolution of the gas
density $\rho_{\rm g}(t)$ in structures at each cosmic time, which
could be written as
\begin{equation}
\dot\rho_{\rm g}=-\frac{{\rm d}^{2}M_{\star}}{{\rm d}V{\rm
d}t}+\frac{{\rm d}^{2}M_{\rm ej}}{{\rm d}V{\rm d}t}+a_{\rm
b}(t)-o(t)\label{rhogas},
\end{equation}
and the metal enrichment history of the IGM could be described by
\begin{eqnarray}
Z_{\rm IGM}(t)={\int_t^{t(z_{\rm ini})} o(t) {\rm d}t\over
\rho_{c}\Omega_{\rm b}-\int_t^{t(z_{\rm ini})}a_{\rm b}(t){\rm
d}t+\int_t^{t(z_{\rm ini})}o(t){\rm d}t},\label{Metallicity}
\end{eqnarray}
which is the mass fraction of metal populated gas to IGM gas. Here
we assume that stars begin to form at the initial refshift of
$z_{\rm ini}=30$. For Pop I/II stars, we derive $\tau_1$ by
comparing the model-predict SFR with the observation from
\cite{Madau14}. We find that $\tau_1=3.8~{\rm Gyr}$ could reproduce
the low-redshift SFR quite well ($z\lesssim4$), and we derive
$f_1=0.83$ by normalizing the local SFR to
$0.016\msun\yr^{-1}\mpc^{-3}$. Therefore, the star formation
efficiency  of Pop I/II stars (or the efficiency of conversion of
baryons in the halo to Pop I/II stars) is $0.83~e^{-(t-t_{\rm
init})/({\rm 3.8 Gyr})}$, which is consistent with the results from
the abundance matching techniques or the gravitational lensing
measurements at $z\lesssim 4$ \citep{Mandelbaum06, Shankar06,
Moster13, Velander14}.

\begin{figure}
\centering\resizebox{0.8\textwidth}{!}{\includegraphics{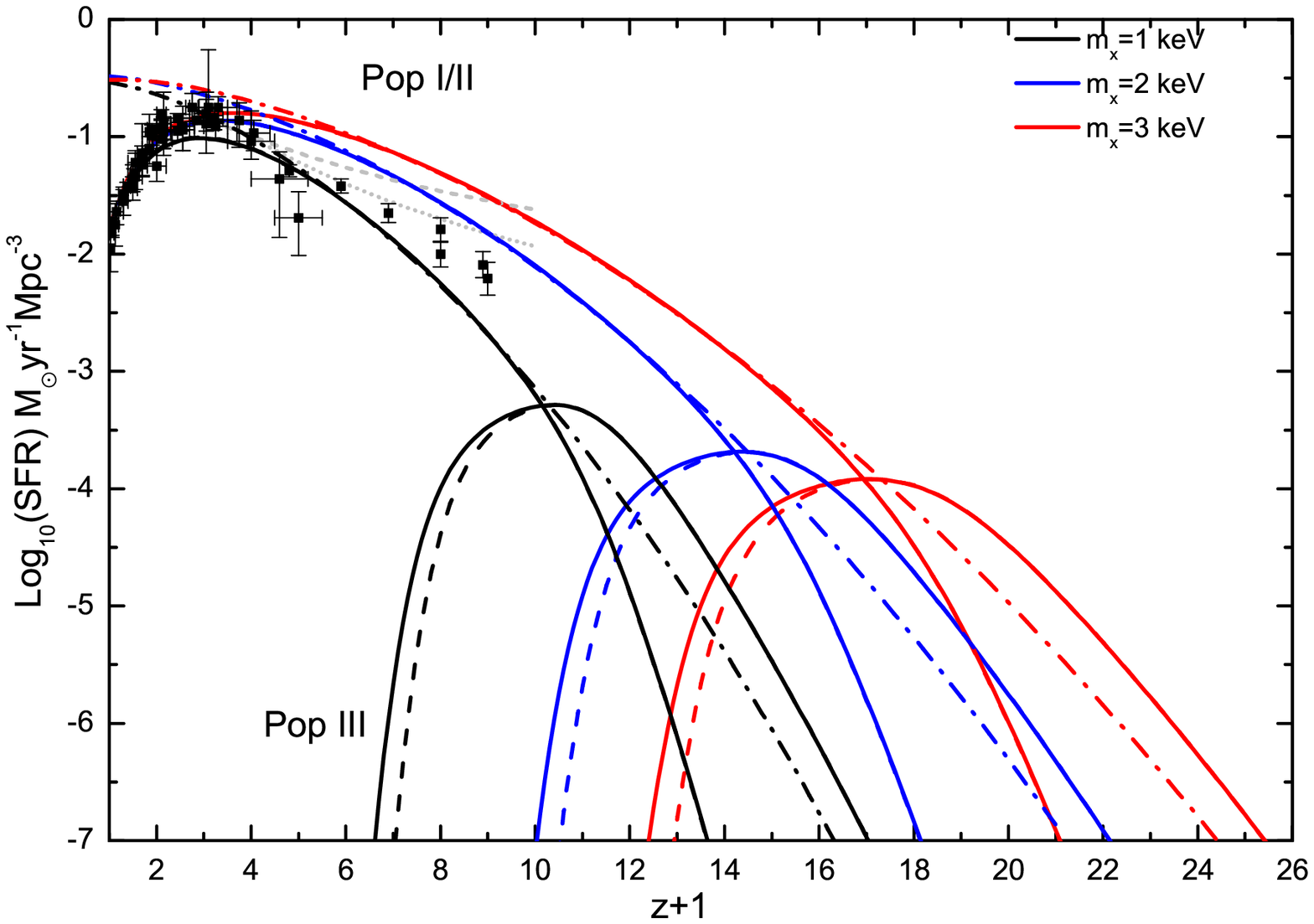}}
\caption{SFRs for Pop I/II and Pop III stars with different WDM
particle masses (from bottom to top): $m_{\rm x}=1 \kev$ (black
lines), $m_{\rm x}=2 \kev$ (blue lines), $m_{\rm x}=3 \kev$ (red
lines). The solid lines correspond to the SIMF and the dashed lines
correspond to the CIMF. The gray dashed and dotted lines are the
SFRs taken from \cite{Robertson15}. The dash-dotted lines are the
SFRs calculated by the relation of SFR$\propto M_{\rm h}^{0.9}$ in a
structure. The observation data is taken from \cite{Madau14}, which
includes the measurements from far-ultraviolet and infrared
luminosity functions of galaxies.}\label{Figure1}
\end{figure}

\begin{figure}
\centering\resizebox{0.8\textwidth}{!}{\includegraphics{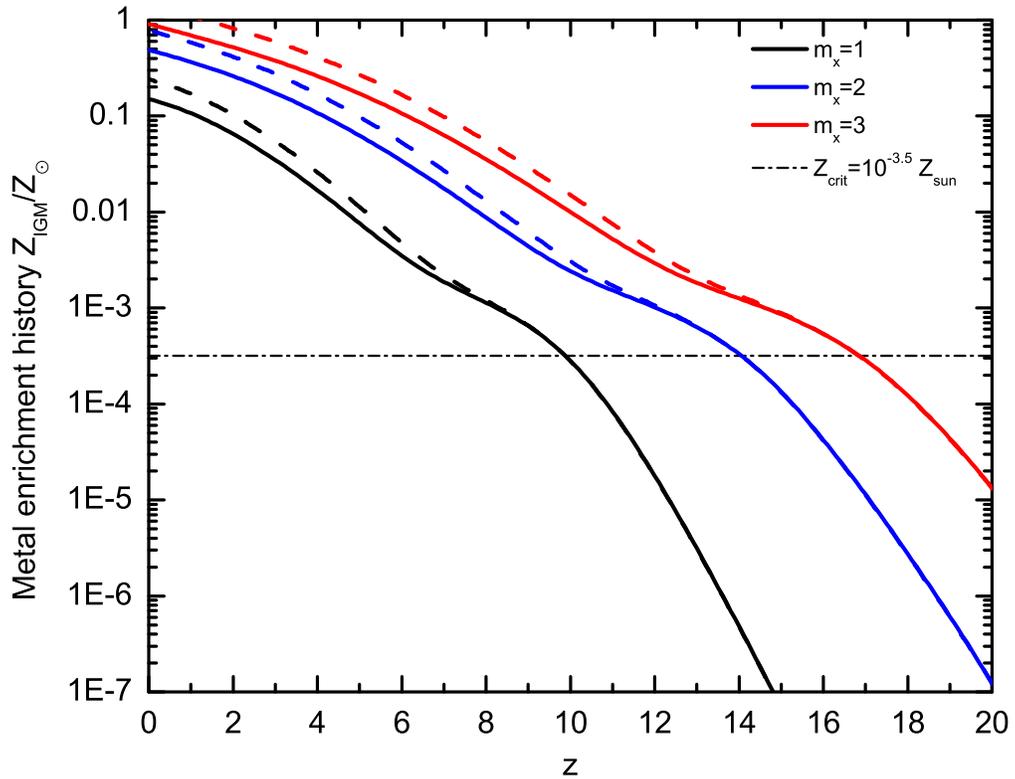}}
\caption{The metal enrichment history of the IGM correspond to the
SFRs in Figure \ref{Figure1}. The solid lines correspond to SIMF,
and the dashed lines correspond to CIMF. The dash-dotted line
corresponds to the critical metallicity of $Z=10^{3.5}Z_\odot$.
}\label{Figure2}
\end{figure}

In Figure \ref{Figure1}, we show the SFRs obtained from the
self-consistency model with different WDM particle masses for Pop
I/II and Pop III stars: $m_{\rm x}=1 \kev$ (black solid lines for
SIMF, black dashed lines for CIMF), $m_{\rm x}=2 \kev$ (blue solid
lines for SIMF, blue dashed lines for CIMF) and $m_{\rm x}=3 \kev$
(red solid lines for SIMF, red dashed lines for CIMF). The observed
SFR is taken from \cite{Madau14}. Our model could reproduce the
observed SFR at low redshift. The high-redshift SFR varies
significantly, and it is sensitive to the WDM particle mass.
Especially for Pop III stars, the peak SFR could be orders of
magnitude different. Comparing with the observations, we could give
a crude constraint on the WDM particle mass. The WDM particle mass
should be larger than $1\kev$, because the model is insufficient to
reproduce the current observation of SFR at redshift $z\gtrsim 5$.
In order to test the validation of our model and the parameters
involved, we compared our SFRs with the one derived from UV
luminosity functions, e.g., SFRs from \cite{Robertson15} (shown as
the gray lines in Figure \ref{Figure1}), and the results are
comparable for $1\kev\lesssim m_{\rm x}\lesssim 3\kev$ at
$z\lesssim10$. On the other hand, we calculated the SFRs by assuming
that the SFR (in units of ${\rm M}_{\odot} \yr^{-1}$) in a structure
is directly connected to the halo mass ($M_{\rm h}$), which is SFR
$\propto M_{\rm h}^\alpha$. We find that a value of $\alpha=0.9$
could give the good predictions for SFRs at $3<z<9$ for all $m_{\rm
x}$ (shown as the dash-dotted lines in Figure \ref{Figure1}), which
is also consistent with results derived from the abundance matching
techniques \citep[e.g.,][]{Shankar06, Moster13, Aversa15}.
Furthermore, the transition from Pop III stars to Pop I/II stars is
quite different, which is mainly determinated by the metal
enrichment history. As shown in Figure \ref{Figure2}, we calculated
the metal enrichment history of IGM for different WDM particle
masses (solid lines for SIMF and dashed lines for CIMF). The
transition occurs at $z=10$ for $m_{\rm x}=1\kev$, $z=14$ for
$m_{\rm x}=2\kev$ and $z=17$ for $m_{\rm x}=3\kev$, which is
consistent with the previous results \citep[e.g.,][]{Yang15}.
Considering that the transition should not occur at a too much high
redshift and the metallicity of IGM should not exceed $Z_{\odot}$
\citep{Daigne06}, it seems that the WDM particle mass should less
than $3\kev$.

\section{Cosmic reionization}

The cosmic reionization history is dependent on the high-redshift
SFR, meanwhile the high-redshift SFR is sensitively dependent on the
WDM particle mass. So the cosmic reionization history could be a
useful tool to probe the WDM particle mass. We assume that the
cosmic reionization is dominated by the high-redshift stars. Using
the SFR derived in the above section, we calculate the rate of
ionizing ultraviolet photons escaping from stars into IGM, which
reads
\begin{eqnarray}
\dot{n}_\gamma(z)=(1+z)^3\left({\Psi(z)_{\rm I/II}\over m_{\rm B}}
N_\gamma^{\rm I/II}f_{\rm esc}^{\rm I/II}+{\Psi(z)_{\rm III}\over
m_{\rm B}} N_\gamma^{\rm III}f_{\rm esc}^{\rm
III}\right)\label{ngamma},
\end{eqnarray}
where $(1+z)^3$ accounts for the conversion of the comoving density
into the proper density, $\Psi(z)$ is the SFR, $m_{\rm B}$ is the
baryon mass, $N_\gamma$ are the number of ionizing UV photons
released per baryon, and $f_{\rm esc}$ are the escape fractions of
these photons from stars into IGM. Here we take the escape fraction
as the constant with $f^{\rm I/II}_{\rm esc}=0.2$, $N_{\gamma}^{\rm
I/II}=4000$, $f^{\rm III}_{\rm esc}=0.7$, and $N_{\gamma}^{\rm
III}=9\times 10^4$ \citep{Greif06}. For another point, many works
have shown that the escape fraction should evolve with redshift.
Following \cite{Hayes11}, the redshift evolution of $f_{\rm esc}$
could be described by
\begin{equation}
f_{\rm esc}(z)=\\
\left\{ \begin{array}{lcl} \left({1+z\over12.1}\right)^{2.57}, &
{\rm if}~
z\leq 11.1,\,\\
1, & {\rm if}~ z> 11.1.\,
\end{array}\right. \label{fescz}
\end{equation}
By defining the volume filling fraction of ionized hydrogen $Q_{\rm
H_{\rm II}}$, we could calculate it from the differential equation
\citep[e.g.,][]{Barkana01, Wang13, Robertson13, Robertson15}.

\begin{eqnarray}
\dot{Q}_{\rm H_{\rm II}}={\dot{n}_{\gamma}(z) \over (1+y)n_{\rm
H}(z)}-\alpha_{B}C(z) (1+y)n_{\rm H}(z) Q_{\rm H_{\rm
II}}\label{reionize},
\end{eqnarray}
where $n_{\rm H}(z)=1.9\times 10^{-7} (1+z)^3~\cm^{-3}$ is the
number density of hydrogen, and
$\alpha_{B}=2.6\times10^{-13}\cm^3\s^{-1}$ is the recombination
coefficient for electron with temperature at about $10^4K$. The
factor $y$ is introduced by considering the ionization of helium,
because the universe at the reionization epoch includes both
hydrogen and helium, whose mass fractions are $X=0.74$ and $Y=0.26$
\citep{Pagel00}, respectively. Here we assume that the helium was
only once ionized, therefore, we derive $y=Y/(4X)\approx0.08$. The
clumping factor of the ionized gas is defined by $C\equiv {\langle
n_{\rm H~_{II}}^2\rangle / \langle n_{\rm H~_{II}}\rangle^2}$=2.9
\citep[e.g.,][]{Pawlik09, Shull12}.

Combing equations (\ref{ngamma}) and (\ref{reionize}), we could give
a numerical solution for $Q_{\rm H_{\rm II}}(z)$ by setting $Q_{\rm
H_{\rm II}}=0$ at the initial refshift of $z_{\rm ini}=30$.
Therefore, the CMB optical depth can be calculated by integrating
the electron density times the Thomson cross section along proper
length as
\begin{eqnarray}
\tau=-(1+y)\sigma_{T} c\int_0^{z_{\rm ini}}n_{\rm H}(z) Q_{\rm
H_{\rm II}}(z){{\rm d}t\over {\rm d}z}{\rm d}z.
\end{eqnarray}
Here, the upper limit of the integral value is $z_{\rm ini}\sim 30$,
because the CMB optical depth could be mainly contributed by the
electrons at relatively low redshift with $z\ll z_{\rm ini}$
\citep{Larson11}.

\begin{figure}
\centering\resizebox{0.8\textwidth}{!}{\includegraphics{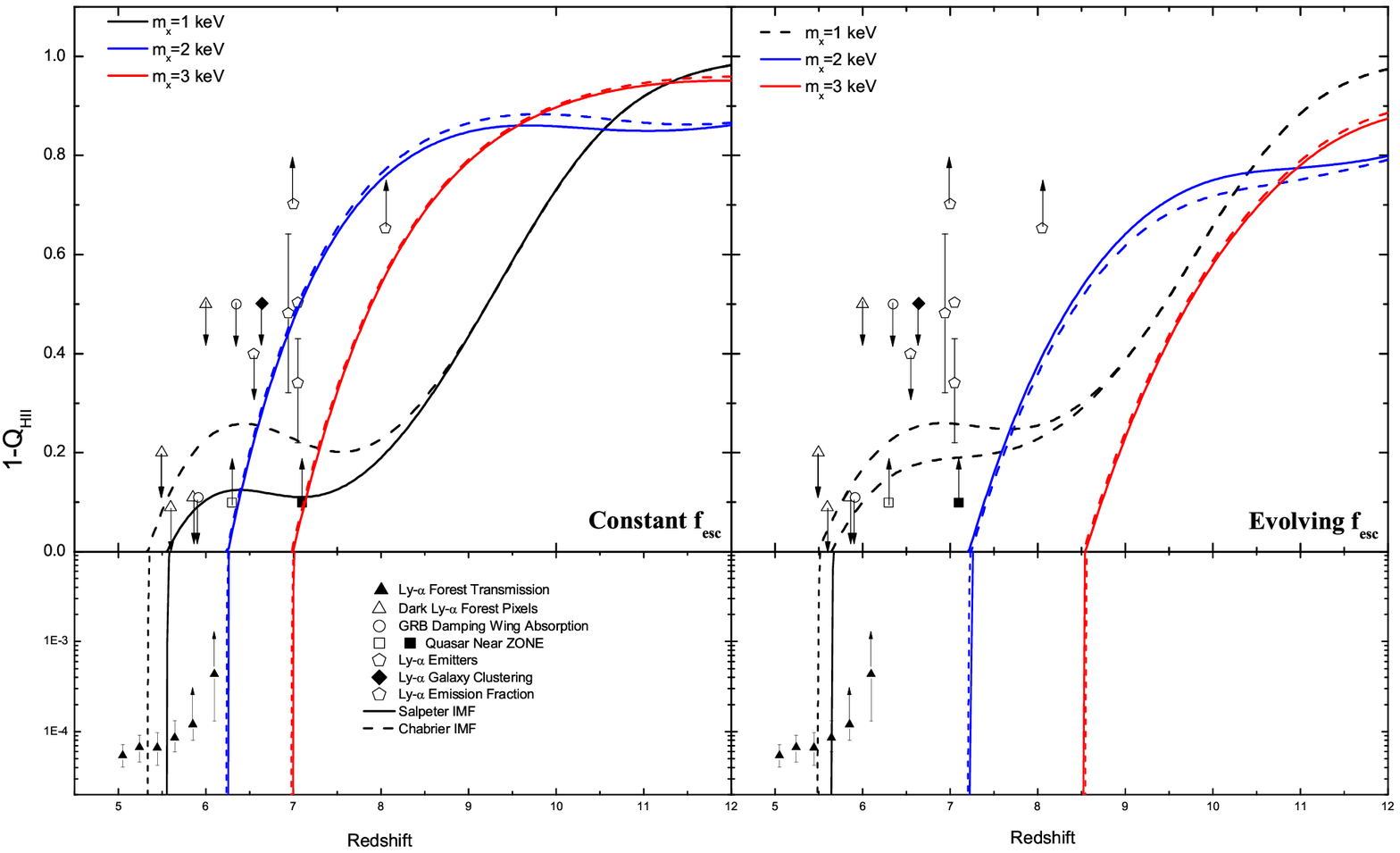}}
\caption{The corresponding neutral fraction of $1-Q_{\rm H_{\rm
II}}$ for the SFRs given in Figure \ref{Figure1}. Left: The neutral
fractions derived by assuming a constant $f_{\rm esc}$. Right: The
neutral fractions derived by assuming an evolving $f_{\rm esc}$ of
equation (\ref{fescz}). Measurements of IGM neutral fractions are
derived from Ly$\alpha$ emitting galaxies \citep{Pentericci14,
Schenker14}, constraints from the Ly$\alpha$ of GRB host galaxies
\citep{Chornock13}, and inferences from dark pixels in Ly$\alpha$
forest measurement \citep{McGreer15}. }\label{Figure3}
\end{figure}

\begin{figure}
\centering\resizebox{0.8\textwidth}{!}{\includegraphics{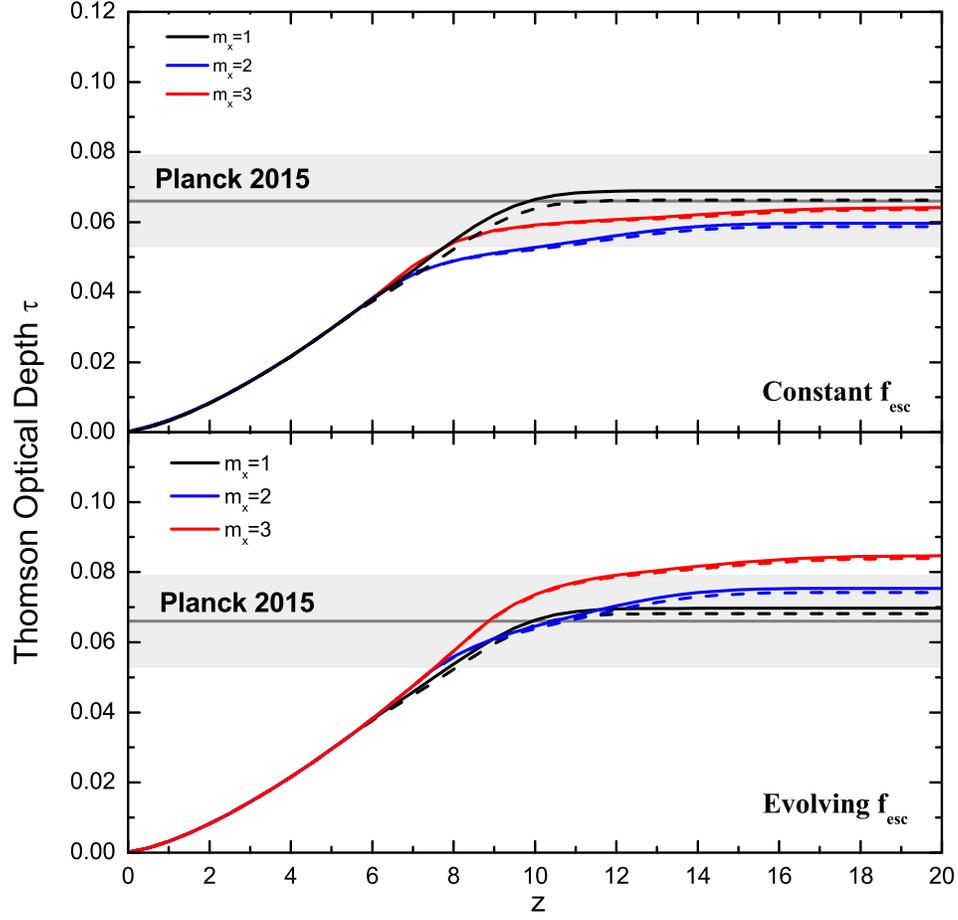}}
\caption{Top: The model-predicted CMB optical depth ($\tau$)
correspond to the reionization of the left panel of Figure
\ref{Figure3}, where a constant escape fraction is assumed. Bottom:
The model-predicted CMB optical depth ($\tau$) with an evolving
escape fraction, which corresponds to the reionization of the right
panel of Figure \ref{Figure3}. The shade region is the CMB optical
depth measured by \emph{Planck} with
$\tau=0.066^{+0.013}_{-0.013}$.}\label{Figure4}
\end{figure}

Observation on the ionization fraction $Q_{\rm H_{\rm II}}$ is
making great progress: the star-forming galaxies showing Ly$\alpha$
emission up to $z\sim 7-8$ \citep{Treu13, Pentericci14, Schenker14};
the Ly$\alpha$ damping wing absorption constrains from GRB host
galaxies \citep{Chornock13}; the number of dark pixels in Ly$\alpha$
forest observation of background quasars \citep{McGreer15}. Most of
the observations give strong evidence that the reionization ending
rapidly near $z\simeq 6$. Figure \ref{Figure3} shows the neutral
fraction of $1-Q_{\rm H_{\rm II}}$ from observations and constrains
from SFRs shown in Figure \ref{Figure1}. The corresponding CMB
optical depth are shown in Figure \ref{Figure4}. Comparing with
these observations, we could give a robust constraint on the SFR,
and hence on WDM particle mass. For constant $f_{\rm esc}$, the
neutral fraction of $1-Q_{\rm H_{\rm II}}$ and the CMB optical depth
$\tau$ are shown in the left panel of Figure \ref{Figure3} and top
panel Figure \ref{Figure4}, respectively. For $m_{\rm x}=1\kev$, the
reionization photons are mainly contributed by Pop III stars at
$z\gtrsim7.5$ and recombination begin to dominate at $6\lesssim
z\lesssim7.5$ until Pop I/II stars dominate the reionization at
$z\lesssim6$. The reionization ended at $z\simeq 5.5$. The neutral
fraction could fit the observation at $z\lesssim7$ but much lower
for $z\gtrsim7$. For $m_{\rm x}=2\kev$, the reionization dominate by
Pop III stars at $z\gtrsim 10$, and Pop I/II stars dominate the
reionization at $z\lesssim10$ with the fully reionized epoch at
$z\simeq6.3$, which fits the observation quite well. For $m_{\rm
x}=3\kev$, Pop III stars contribute little photons to the
reionization at low redshifts and Pop I/II stars dominate the
reionization at $z\lesssim 12$, the reionization ended at $z\simeq
7$. However, this is a little farfetched to the observation.
Therefore, it produces too much reionization photons at redshift
$z\gtrsim 7$ for $m_{\rm x}\lesssim1 \kev$ or $m_{\rm x}\gtrsim3
\kev$, which can not fit the observations. For the CMB optical depth
of $\tau$, they all located in the range of the measurement from
\emph{Planck} \citep{Planck Collaboration15}. For the evolving
$f_{\rm esc}$, the natural fraction of $1-Q_{\rm H_{\rm II}}$ and
$\tau$ are shown in the right panel of Figure \ref{Figure3} and
bottom panel Figure \ref{Figure4}, respectively. The reionization
process is faster than that of the constant $f_{\rm esc}$, because
more reionization photons are escaped into IGM. The results show
that only a narrow range of $1\kev<m_{\rm x}< 2\kev$ could fit the
observations of $1-Q_{\rm H_{\rm II}}$ and $\tau$. For $m_{\rm
x}=3\kev$, neither the neutral fraction nor the CMB optical depth
could fit the observations.

\section{SGWB from BH `ringing'}

In this section, we will calculate the SGWB from `ringing' BHs,
which are quite dependent on the SFRs of Pop I/II and Pop III stars.
The total GW flux received on earth could be written as
\begin{equation}
F_\nu(\nu_{\rm obs})=\int {1 \over 4\pi d_{\rm L}^2} {{\rm d}E_{\rm
GW}\over {\rm d}\nu} {{\rm d}\nu \over {\rm d}\nu_{\rm obs}} \Psi(z)
\Phi(m){\rm d} m{\rm d}V,
\end{equation}
where $d_{\rm L}$ is the luminosity distance, ${{\rm d}E_{\rm GW}/
{\rm d}\nu}$ is the GW energy spectrum of an individual source,
$\Psi(z)$ is the SFR, $\Phi(m)$ is the IMF as shown in equation
(\ref{Salpeter}) and (\ref{Chabrier}). ${\rm d}V$ is the comoving
volume element. In the above equation, the observed GW energy flux
per unit frequency for an individual source is
\begin{equation}
f_\nu(\nu_{\rm obs})= {1 \over 4\pi d_{\rm L}^2} {{\rm d}E_{\rm
GW}\over {\rm d}\nu} {{\rm d}\nu \over {\rm d}\nu_{\rm obs}},
\end{equation}
which could also be written as \citep{Carr80}
\begin{equation}
f_\nu(\nu_{\rm obs})={\pi c^3\over 2G}h^2_{\rm BH},
\end{equation}
where $h_{\rm BH}$ is the dimensionless GW amplitude produced by a
star collapses into a BH. The total GW flux received on earth could
also be written as
\begin{equation}
F_\nu(\nu_{\rm obs})={\pi c^3\over 2G}h^2_{\rm BG}\nu_{\rm obs}.
\end{equation}
Combing with the above equations, we obtain
\begin{equation}
h^2_{\rm BG}={1\over \nu_{\rm obs}}\int h^2_{\rm BH}\Psi(z)
\Phi(m){\rm d} m{\rm d}V.
\end{equation}
Following \cite{Thorne87}, the dimensionless amplitude for a star
collapses into to a BH with mass $m_r$ is given by
\begin{equation}
h_{\rm BH}\simeq7.4\times 10^{-20}\epsilon_{\rm
GW}^{1/2}\left({m_{\rm r}\over \msun}\right)\left({ d_{\rm L}\over
1\mpc}\right)^{-1},
\end{equation}
where $\epsilon_{\rm GW}\lesssim 7\times 10^{-4}$ is the GW
radiation efficiency. The corresponding GW frequency in the observer
frame is
\begin{equation}
\nu_{\rm obs}\simeq 1.3\times 10^4
 {\rm Hz} \left({\msun\over m_{\rm r}}\right)(1+z)^{-1},
 \end{equation}
where $(1+z)^{-1}$ accounts for the redshift effect. It is obvious
that the observed GW frequency is quite dependent on the BH mass.
The maximum BH remnant for Pop I/II stars is $43.3\msun$, whereas
$500\msun$ for Pop III stars \citep{Pereira10}. Therefore, the
minimum GW frequency would be 12 times lower for Pop III BHs
(considering the no time delay between the formation of Pop I/II and
Pop III stars).

Usually, the SGWB is described by the dimensionless energy density
parameter $\Omega_{\rm GW}(\nu_{\rm obs})$, which is the present GW
energy density per logarithmic frequency interval divided by the
critical energy density of the present universe ($\rho_{\rm c}c^2$)
\citep{Phinney01}
\begin{eqnarray}
\Omega_{\rm GW}(\nu_{\rm obs})={1\over \rho_c c^2}{{\rm d}\rho_{\rm
gw}\over{\rm d~ln}\nu_{\rm obs}},
\end{eqnarray}
where $\rho_{\rm gw}$ is the GW energy density, and $\rho_{\rm
c}={3H_0^2/ 8 \pi G}$ is the critical density of the universe. For
the astrophysical origin of the SGWBs, $\Omega_{\rm GW}(\nu_{\rm
obs})$ could be written as
\begin{eqnarray}
\Omega_{\rm GW}(\nu_{\rm obs})={\nu_{\rm obs}\over \rho_{\rm c} c^3}
F_{\nu_{\rm obs}}(\nu_{\rm obs})={4\pi^2\over 3H_0^2}\nu_{\rm obs}^2
h_{\rm BG}^2.
\end{eqnarray}
To evaluate the detectability of the GW signal, we also calculated
the signal to nose (SNR) for a pair interferometers
\citep[e.g.,][]{Christensen92, Flanagan93, de Araujo05, Regimbau06},
\begin{eqnarray}
({\rm S/N})^2={9 H_0^4 \over 50\pi^4}T\int_0^\infty {\rm d}\nu
{\gamma^{2}(\nu)\Omega_{\rm GW}^2(\nu)\over \nu^6 S_{\rm
h}^{(1)}S_{\rm h}^{(2)}},
\end{eqnarray}
where $T=1 \yr$ is the observation time period, $S_{\rm h}^{(i)}$ is
the spectral noise density, and $\gamma(\nu)$ is the overlap
reduction function. For a simple consideration, we assume  $S_{\rm
h}^{(1)}= S_{\rm h}^{(2)}$ and $\gamma=1$ in our calculation.

\begin{figure}
\centering\resizebox{0.8\textwidth}{!}{\includegraphics{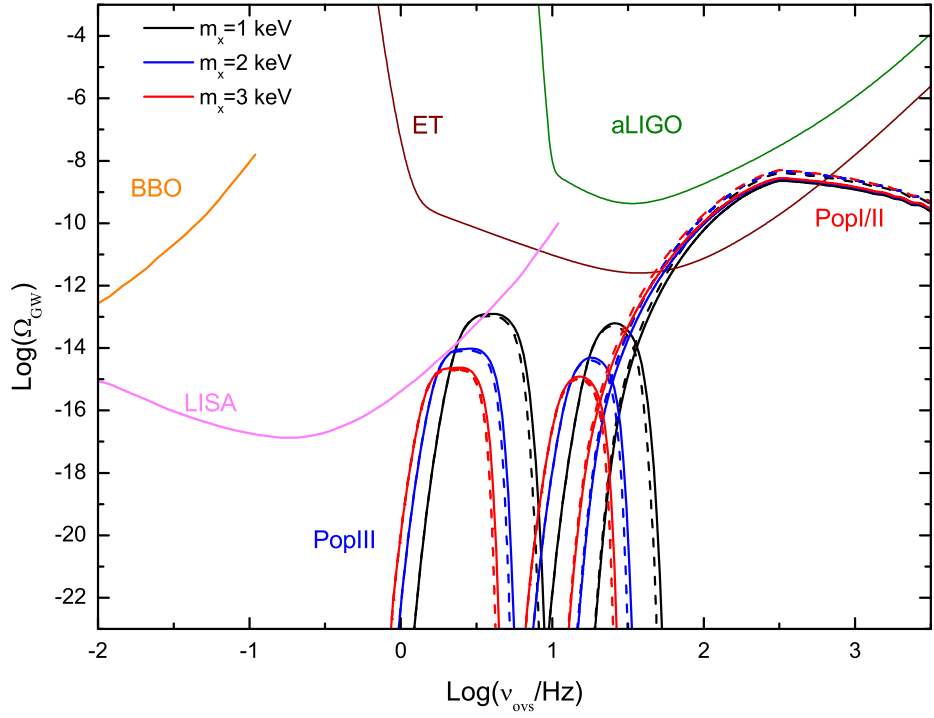}}
\caption{The stochastic gravitational wave background from Pop I/II
and Pop III black holes, which correspond to the SFRs in Figure
\ref{Figure1}. The detection thresholds of advanced LIGO, ET, LISA,
and BBO are labeled with different colors, where one year of
observation is assumed.}\label{Figure5}
\end{figure}

Figure \ref{Figure5} shows the SGWBs from Pop I/II and Pop III BHs.
The sensitivity of GW detectors are also shown, where one year of
observation are assumed \citep{Abadie10, Hild11, Thrane13}. In our
calculation, three WDM particle masses with $m_{\rm x}=1\kev$,
$m_{\rm x}=2\kev$ and $m_{\rm x}=3\kev$ are considered. For Pop I/II
BHs, the SGWB peaks at $\nu=316 {\rm Hz}$ with the amplitude
$\Omega_{\rm GW}=2.8\times 10^{-9}~(5.0\times 10^{-9})$ for SIMF
(CIMF), which is above the delectability of ET. The SGWBs from Pop
I/II BHs are nearly the same for these three WDM particle masses,
with only little difference at lower frequencies. We also calculated
the the SNR, which are $\rm SNR=27$~(47.7), $34.2~(60.5)$ and
$37.7~(66.5)$ for $m_{\rm x}=1\kev$, $m_{\rm x}=2\kev$ and $m_{\rm
x}=3\kev$ for ET with SIMF (CIMF), respectively. Therefore, it is
impossible to constrain the WDM particle mass with SGWB from Pop
I/II BHs. For Pop III BHs, the SGWB shifts to lower frequencies, and
the amplitudes are much lower than Pop I/II BHs. The lack of GW
radiation at frequency $\sim 10~ {\rm Hz}$ is because stars within
mass range of $140-260 \msun$ explode as pair-instability supernovae
without leaving BHs. Moreover, the SGWB amplitude with $m_{\rm
x}=1\kev$ is nearly an order of magnitude higher comparing with
$m_{\rm x}=3\kev$, and the peaks shift to lower frequencies. The
most interesting result is that the SGWB from Pop III BHs with
$m_{\rm x}=1\kev$ is detectable for LISA, and the SNR is 1.76 (1.7)
for SIMF (CIMF). However, it could not be detected for $m_{\rm
x}=3\kev$, because the SNR is only $0.35 (0.33)$. On the other hand,
if LISA detectes the SGWB from Pop III BHs, it will be acceptable
for $m_{\rm x}=1\kev$. However, if LISA will not detect the SGWB
from Pop III BHs, the value of $m_{\rm x}=1\kev$ could be excluded.
Therefore, we suggest that the SGWB from Pop III BHs could be
another useful tool to constrain the WDM particle mass.

\section{Conclusion and Discussion}
Although the CDM paradigm has great success in explaining the large
scale structure of the universe, it still have some problems on
small scales. An alternatively WDM paradigm could ease these
problems by employing the $\kev$ WDM particles. In this paper, we
calculate the SFRs of Pop I/II and Pop III stars in the framework of
WDM paradigm. By using a self-consistent method, we reproduce the
SFR at low redshift. We find that the high-redshift SFR is
sensitively dependent on the WDM particle mass, especially for Pop
III stars. By comparing the model-predicted SFR with the
observation, we constrain the WDM particle mass with $m_{\rm
x}>1\kev$, shown as the black lines in Figure \ref{Figure1}. We also
calculated the metal enrichment history of IGM, and the transition
from Pop III to Pop I/II stars is consistent with the previous
results \citep[e.g.,][]{Yang15}, e.g., from $z\sim 10-17$ for
$m_{\rm x} \sim 1-3\kev$. By considering that the metallicity of IGM
does not exceed $Z_{\sun}$, the WDM particle mass should less than
$3\kev$.

Combing with the CMB optical depth from \emph{Planck} with
$\tau=0.066^{+0.013}_{-0.013}$ and the ionization fraction $Q_{\rm
H_{\rm II}}$ from recent observations, we found that the the WDM
particle mass should in the range of $1\kev \lesssim m_{\rm
x}\lesssim3\kev$, where we have assumed a constant escape fraction
of ionizing photons \citep[e.g.,][]{Schultz14, Dayal15a}. However,
many works suggest that the escape fraction should be redshift
dependent \citep[e.g.,][]{Siana10, Blanc11, Hayes11, Kuhlen12,
Dijkstra14}. By considering an evolving escape fraction, we found a
more tight constraint $1\kev< m_{\rm x}<2\kev$.

Finally, recent observation of GW150914 and GW151226 inspires a
great interest in the field of GW. Therefore, we calculated the
SGWBs form Pop I/II and Pop III BHs. Our results show that the SGWB
from Pop I/II BHs is not sensitive to the WDM particle mass, and it
could be detected by the ET telescope. However, it is impossible to
constrain the WDM particle mass by the SGWB from Pop I/II BHs,
because they show little difference for different $m_{\rm x}$. For
Pop III stars, the SGWB is quite dependent on the WMD particle mass.
The peak SGWB amplitude with $m_{x}=1\kev$ is an order of magnitude
higher than $m_{x}=3\kev$. The corresponding SNR are $1.76~(1.7)$
and 0.33 (0.35) for SIMF (CIMF), respectively, which is
distinguishable for LISA. Moreover, the SGWBs are derived by
assuming a maximum GW generation efficiency of $\epsilon_{\rm
GW}=7\times 10^{-4}$. Combing with the ET observation of SGWB from
Pop I/II BHs, we could give a constraint on $\epsilon_{\rm GW}$.
Therefore, a further constraint of $m_{\rm x}$ (or Pop III SFR)
could be given by the observation of LISA. On the other hand, the
SGWB from Pop III BHs is also quite dependent on the star formation
efficiency of $f_2$. Therefore, a lower efficiency of $f_2$ will
make it hard to constrain the WDM particle mass by the observation
of LISA. Anyway, the large difference of SGWBs from Pop III BHs for
different $m_{\rm x}$ will make it possible to constrain the WDM
particle mass in future.

\section*{Acknowledgements }
We thank the anonymous referee for valuable comments and
suggestions. This work is supported by the National Basic Research
Program of China (973 Program, grant No. 2014CB845800) and the
National Natural Science Foundation of China (grants 11422325 and
11373022), the Excellent Youth Foundation of Jiangsu Province
(BK20140016), and Jiangsu Planned Projects for Postdoctoral Research
Funds. K.S.C. is supported by the CRF grants of the Government of
the Hong Kong SAR under HUKST4/CRF/13G.

\end{document}